\documentclass[aps,reprint,pra,longbibliography]{revtex4-2}
\usepackage{amssymb}
\usepackage{amsmath}
\usepackage{dcolumn}
\usepackage{bm}
\usepackage{graphicx}
\usepackage{mathrsfs}
\usepackage[colorlinks,linkcolor=blue,anchorcolor=blue,citecolor=blue,urlcolor=blue]{hyperref}
\usepackage{color}

\begin{document}

\title{Intra-band entanglement-assisted cavity electro-optic quantum
transducer}
\author{Yu-Bo Hou}
\thanks{Yu-Bo Hou and Rui-Zhe You contributed equally to this work.}
\author{Rui-Zhe You}
\thanks{Yu-Bo Hou and Rui-Zhe You contributed equally to this work.}
\author{Di-Jia Zhang}
\author{Pengbo Li}
\author{Changchun Zhong}
\email{zhong.changchun@xjtu.edu.cn}

\address{
MOE Key Laboratory for Non-equilibrium Synthesis
and Modulation of Condensed Matter, Shaanxi Province Key Laboratory
of Quantum Information and Quantum Optoelectronic Devices, School of
Physics, Xi’an Jiaotong University, Xi’an 710049, China}

\begin{abstract}
Quantum transduction is a key technology for connecting different quantum
technologies across varied frequencies. However, it remains a major challenge to overcome the high
threshold for achieving positive capacity of
traditional quantum transduction channels. Recently, an
entanglement-assisted transducer was proposed based on a cavity-optic system
[Opt. Quantum 2, 475 (2024)], where a modified bosonic loss channel was
obtained, and the transduction efficiency can be enhanced by properly tuning
the squeezing parameters. In this paper, we further identify three types of
quantum channels enabled by this design, offering additional options for
implementing the proposed transduction schemes. Compared to the transducers
without entanglement assistance, the scheme also shows a great enhancement
in the conversion bandwidth for achieving high quantum capacity, further
increasing its value in practical applications.
\end{abstract}

\maketitle

\section{Introduction}

Quantum transduction refers to the conversion of quantum information between
different physical platforms, typically between microwave and optical
photons \cite{1,2,9}. The two systems differ widely in the energy scale
and involve fundamentally different interaction mechanisms. The microwave
quantum bits, based on superconducting circuits and other quantum
processors, offer excellent coherence and scalability but lack intrinsic
optical transitions \cite{5,6,7}, while the optical photon is ideal
information carriers for long distance quantum communication. \cite{10,11,12}%
. Quantum transduction bridges this gap by enabling coherent coupling
between superconducting qubits and optical photons, which are essential for
constructing large-scale quantum networks and distributed quantum
architectures \cite{13,15,16,17,14,zhong}.

In the past decades, significant progress has been made in quantum
transduction based on various physical platforms, including electro-optics
\cite{21,24,26,29,32,36,68,70}, electro-optomechanics \cite%
{20,27,28,38,39,77,79,74,75,76,78,80,81,82}, piezo-optomechanics \cite%
{31,33,61}, quantum magnonics \cite{22,25,62,63,73}, rare-earth-ions \cite%
{34,66,67,86} and atoms \cite{30,35,64,65,83,84,85}. Theoretically, all quantum
transduction systems can be regarded as a quantum channel. To reliably
transmit encoded quantum information, a quantum channel must have a positive
quantum capacity. This requirement indicates that a quantum transduction
channel, generally modeled as a bosonic loss channel, must have both high
channel transmissivity and low added noise \cite{40}. However, the
traditional direct quantum transducer (DQT), which linearly converts photons
between different frequencies, faces significant challenges in reaching the
positive quantum capacity threshold due to technological constraints, such
as limited interaction strength and excessive thermal noises \cite%
{41,42,43,44,55,56,57}. In order to enhance the performance of transduction channels, numerous approaches have been developed
, such as
entanglement based transduction \cite{46,47,48,49,50,58,59,zhong2}, adaptive
feedforward control \cite{51,52}, single mode squeezing enhanced transducer
\cite{45}.

In Ref. \cite{87}, Haowei discussed a new scheme of an entanglement-assisted
transducer based on a cavity electro-optic (EO) system. Specifically, for
optical to microwave transduction, the scheme first entangles an ancilla with
a probe in the microwave domain through a two-mode squeezer. The probe
output, along with an optical encoding signal, is then sent into the EO
system. The microwave output signal undergoes anti-squeezing with the
ancilla mode using a second squeezer. It is shown that this process defines
a new thermal loss channel whose quantum transduction capacity can be
greatly enhanced. In this paper, we show the process actually induce more
general transduction channels, namely random displacement, generalized
thermal loss and thermal amplification channels. We perform a detailed
analysis of the three types of transduction channels by adjusting the
squeezing strengths of the two squeezers. The quantum capacities of
different transduction channels are quantified in much wide parameter space,
greatly expanding the potential scope of quantum transduction applications.
Furthermore, we compare the cases with and without entanglement assistance
under the non-resonant condition, demonstrating a significant improvement in
quantum transduction bandwidth. These findings underscore the advantages of
this scheme, paving the way of realizing a high bandwidth quantum transducer
potentially in the near term.

\section{Quantum Transducer Based on Cavity Electro-Optic System}

\begin{figure}[tbh]
\centering
\includegraphics[width=\columnwidth]{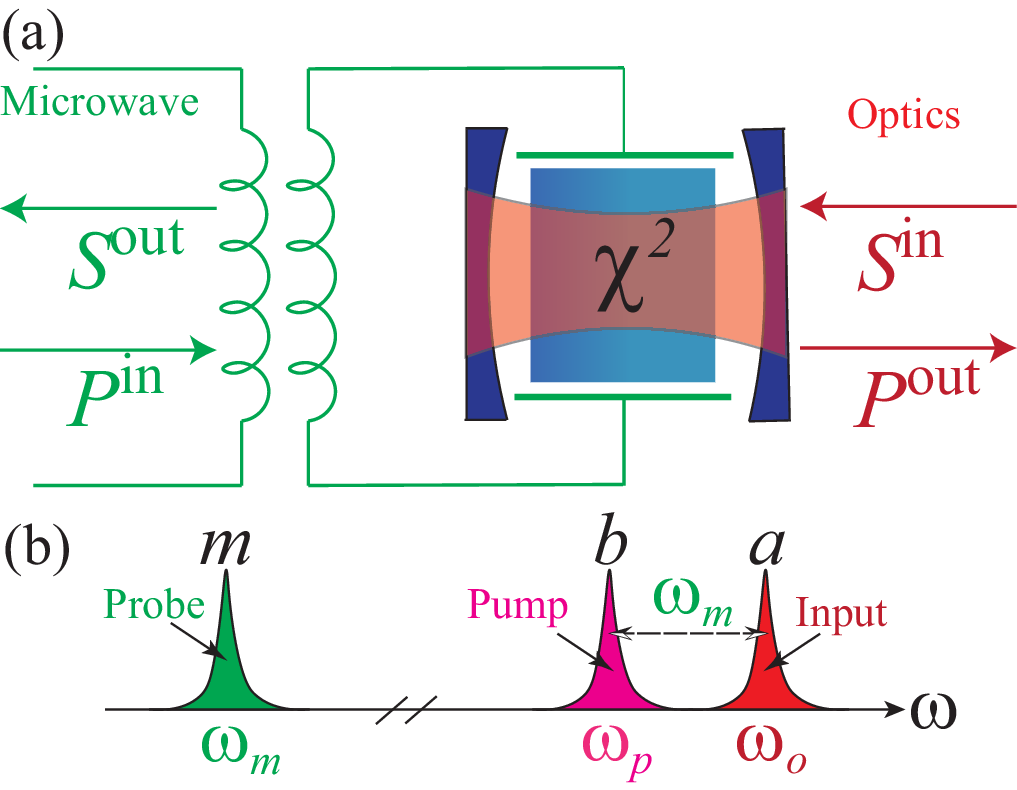}
\caption{(Color online) (a) Schematic diagram of the EO transducer. It
comprises a superconducting resonator integrated with an optical cavity,
consisting of material which features Pockels nonlinearity ($\protect\chi %
^{2}$). The microwave (green) probe signal and the optical (red) input
encoding signal are denoted as $P^{\mathrm{in}}$ and $S^{\mathrm{in}}$,
respectively. The corresponding outputs after the transduction through the
cavity EO system are denoted as $P^{\mathrm{out}}$ and $S^{\mathrm{out}}$.
(b) Three-wave mixing structure based on the Pockels effect. The
lower-frequency optical mode $b$ (pink) is driven by a laser pump to
generate a beam splitter interaction between the microwave mode $m$ (green)
and the optical mode $a$ (red), with the corresponding frequencies $\protect%
\omega _{m}$ and $\protect\omega _{o}$.}
\label{f1}
\end{figure}

We begin by analyzing the quantum transducer based on the superconducting
cavity EO system, as shown in Fig. \ref{f1}(a). The optical cavity, made of
material with Pockels nonliearity, is placed inside the capacitor of an LC
microwave resonator. The electric field generated by the resonator alters
the refractive index of the material in the optical cavity, thereby
modulating its optical resonant frequency. Conversely, the optical fields
within the cavity can induce a microwave field through optical rectification
in the Pockels material, enabling bidirectional interaction between the
optical and microwave domains \cite{71,72}. Moreover, the material's
nonlinear properties are utilized to enable the interaction between two
specific optical modes and a microwave mode through frequency-matched
three-wave mixing \cite{21}, as shown in Fig. \ref{f1}(b). By strongly
pumping the lower-frequency optical mode $b$, with a frequency $\omega _{p}$%
, a beam splitter interaction can be generated between the microwave mode
and the higher-frequency optical mode. In the interaction picture, the total
Hamiltonian is given as (let $\hbar =1$ hereafter)
\begin{equation}
H=-g(a^{\dagger }m+a m^{\dagger }),
\end{equation}
where $a$ and $m$ denote the annihilation operators for the optical and
microwave modes, respectively. Here, the frequency-matching condition $%
\omega _{o}-\omega _{p}=\omega _{m}$ is satisfied, where $\omega_o$ ($\omega_m$) is the optical (microwave) resonant frequency. $g$ is the
laser-enhanced coupling strength.

Now, we consider an optical input signal $S^{\mathrm{in}}$ and a microwave
probe $P^{\mathrm{in}}$ in the EO system, and define $\varepsilon _{X}$ as
the fluctuation operator associated with the $X$ port, satisfying the
commutation relation $[\varepsilon _{X}(t),\varepsilon _{X}^{\dag
}(t^{\prime })]=\delta (t-t^{\prime })$. The system dynamics are governed by
the quantum Langevin equations (QLE), which are given by%
\begin{equation*}
\dot{m}=iga-\frac{\kappa _{m}}{2}m+\sqrt{\kappa _{m,c}}\varepsilon _{P^{%
\mathrm{in}}}+\sqrt{\kappa _{m,i}}\varepsilon _{m}\text{,}
\end{equation*}%
\begin{equation}
\dot{a}=igm-\frac{\kappa _{a}}{2}a+\sqrt{\kappa _{a,c}}\varepsilon _{S^{%
\mathrm{in}}}+\sqrt{\kappa _{a,i}}\varepsilon _{a}\text{.}  \label{e2}
\end{equation}%
Here, the total loss rate of the microwave (optical) mode is defined as $%
\kappa _{m(a)}\equiv \kappa _{m(a),c}+\kappa _{m(a),i}$, with the coupling
and intrinsic loss rates $\kappa _{m(a),c}$ and $\kappa _{m(a),i}$,
respectively. Moreover, $\varepsilon _{m}$ and $\varepsilon _{a}$ are the
quantum noise operators, which obey the correlation functions $\langle
\varepsilon _{m}^{\dag }(t)\varepsilon _{m}(t^{\prime })\rangle =N_{m}\delta
(t-t^{\prime })$ and $\langle \varepsilon _{a}^{\dag }(t)\varepsilon
_{a}(t^{\prime })\rangle =N_{a}\delta (t-t^{\prime })$, respectively. The
mean thermal photon excitation number is given by $N_{m(a)}=[\exp (\hbar
\omega _{m(o)}/k_{B}T)-1]^{-1}$, with the Boltzmann constant $k_{B}$ and the
bath temperature $T$. Notably, the thermal effects in the optical frequency
range can be safely disregarded, because even at room temperature, the
thermal photon number $N_{a}\sim 10^{-22}$ at $\omega _{o}\sim 300$ THz is
negligible.

Then applying the Fourier transform $f(t)=\frac{1}{2\pi }\int_{-\infty
}^{+\infty }f(\omega )e^{-i\omega t}d\omega $, the QLEs in Eq. (\ref{e2})
are transformed into the frequency domain as%
\begin{equation}
\vec{V}=M_{1}\vec{V}+M_{2}\vec{V}_{\mathrm{in}}+M_{3}\vec{V}_{i}\text{,}
\end{equation}%
where $V=[m\left( \omega \right) ,a\left( \omega \right) ]^{T}$ is the mode
vector, while $\vec{V}_{\mathrm{in}}=[\varepsilon _{P^{\mathrm{in}}}(\omega
),\varepsilon _{S^{\mathrm{in}}}(\omega )]^{T}$ and $\vec{V}%
_{i}=[\varepsilon _{m}(\omega ),\varepsilon _{a}(\omega )]^{T}$ are the
input and noise vectors, respectively. Here, the corresponding coefficient
matrices are given by%
\begin{equation}
M_{1}\mathcal{=}\left[
\begin{array}{cc}
0 & \frac{ig}{-i\omega +\frac{\kappa _{m}}{2}} \\
\frac{ig}{-i\omega +\frac{\kappa _{a}}{2}} & 0%
\end{array}%
\right] \text{,}
\end{equation}%
\begin{equation}
M_{2}\mathcal{=}\left[
\begin{array}{cc}
\frac{\sqrt{\kappa _{m,c}}}{-i\omega +\frac{\kappa _{m}}{2}} & 0 \\
0 & \frac{\sqrt{\kappa _{a,c}}}{-i\omega +\frac{\kappa _{a}}{2}}%
\end{array}%
\right] \text{,}
\end{equation}%
and%
\begin{equation}
M_{3}\mathcal{=}\left[
\begin{array}{cc}
\frac{\sqrt{\kappa _{m,i}}}{-i\omega +\frac{\kappa _{m}}{2}} & 0 \\
0 & \frac{\sqrt{\kappa _{a,i}}}{-i\omega +\frac{\kappa _{a}}{2}}%
\end{array}%
\right] \text{.}
\end{equation}
Combining with the standard input-output relation $\varepsilon _{S^{\mathrm{out}}}(\omega )=\sqrt{\kappa _{m,c}}m(\omega )-\varepsilon _{P^{\mathrm{in}}}(\omega )$, we can figure out%
\begin{equation}
\varepsilon _{S^{\mathrm{out}}}(\omega )=\sqrt{\eta }\varepsilon _{S^{%
\mathrm{in}}}(\omega )+\sqrt{\kappa _{P}}\varepsilon _{P^{\mathrm{in}%
}}(\omega )+\sqrt{\kappa _{E}}\varepsilon _{E}(\omega )\text{,}  \label{e4}
\end{equation}%
with the transduction efficiency spectrum%
\begin{equation}
\eta (\omega )=\left\vert \frac{ig\sqrt{\kappa _{m,c}\kappa _{a,c}}}{%
(-i\omega +\frac{\kappa _{m}}{2})(-i\omega +\frac{\kappa _{a}}{2})+g^{2}}%
\right\vert ^{2}\text{,}  \label{eta_intrinsic_eo}
\end{equation}%
and the probe transmissivity spectrum%
\begin{eqnarray}
&&\kappa _{P}(\omega ) \\
&=&\left\vert \frac{\omega ^{2}+i(\frac{\kappa _{m}+\kappa _{a}}{2}-\kappa
_{m,c})\omega +\frac{\kappa _{a}\kappa _{m,c}}{2}-\frac{\kappa _{m}\kappa
_{a}}{4}-g^{2}}{(-i\omega +\frac{\kappa _{m}}{2})(-i\omega +\frac{\kappa _{a}%
}{2})+g^{2}}\right\vert ^{2}\text{.}  \notag
\end{eqnarray}%
Additionally, $\kappa _{E}$ is the transmissivity of the loss port $E$,
which comes from the intrinsic loss of the two modes. These transmissivities
satisfy the normalization condition $\eta +\kappa _{P}+\kappa _{E}=1$.

\begin{figure}[tbh]
\centering \includegraphics[width=8cm]{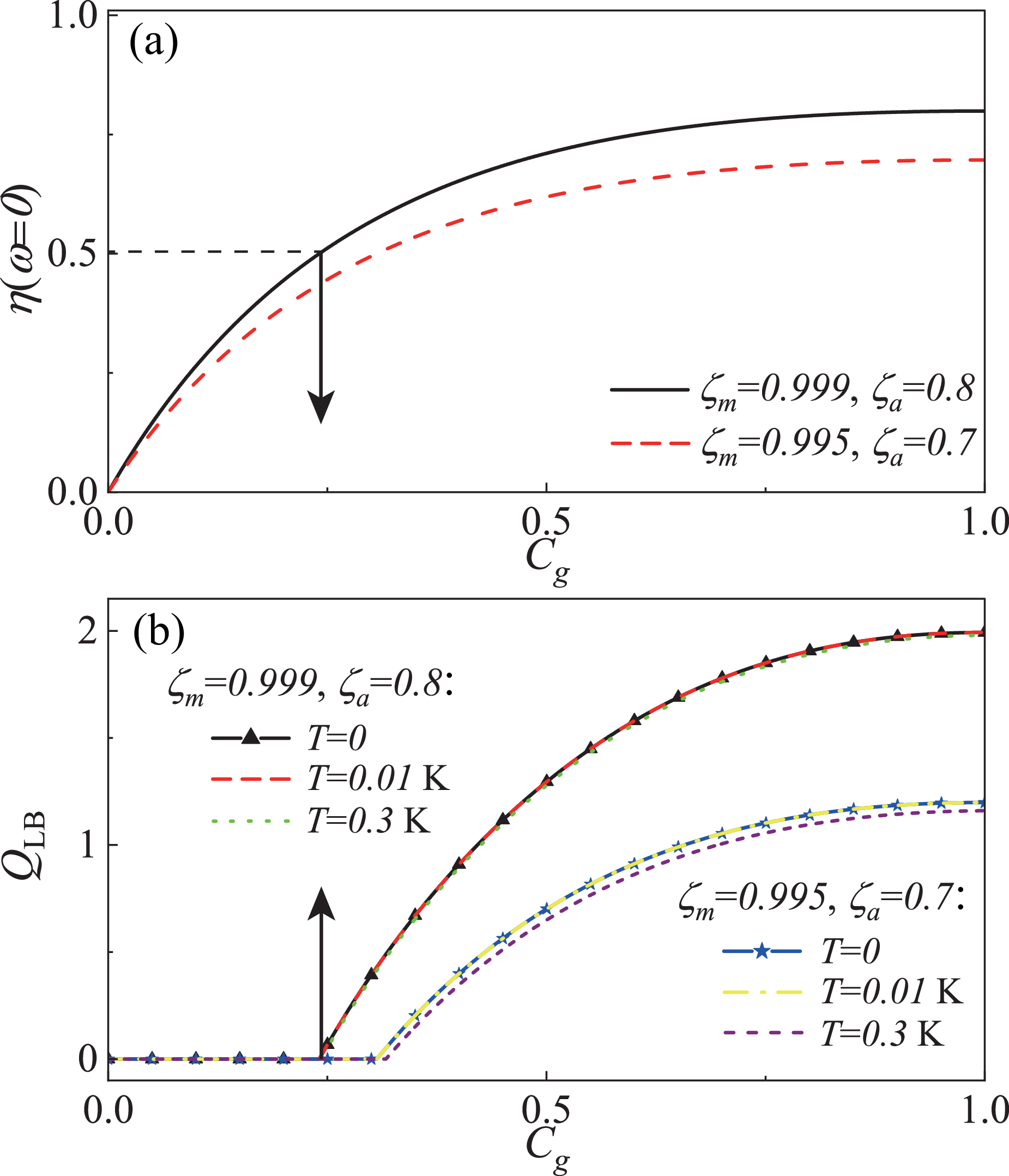}
\caption{(Color online) Quantum transduction in EO system. (a) Transduction
efficiency $\protect\eta $ and (b) lower bound of the quantum capacity $Q_{%
\mathrm{LB}}$ as a function of the cooperativity $C_{g}$ in the resonant
case. The frequency of the microwave mode is chosen as $\protect\omega _{m}=2%
\protect\pi \times 10$ GHz.}
\label{f3}
\end{figure}

The system forms a quantum transduction channel. Take the system on
reasonance for example, it gives a single-mode bosonic loss channel with the
transduction efficiency
\begin{equation}
\eta (\omega =0)=\frac{4C_{g}}{(1+C_{g})^{2}}\zeta _{m}\zeta _{a}\text{,}
\label{e7}
\end{equation}%
and the probe transmissivity%
\begin{equation}
\kappa _{P}(\omega =0)=(\frac{2\zeta _{m}}{1+C_{g}}-1)^{2}\text{.}
\end{equation}%
Here, we define the coupling ratio of the microwave (optical) mode as $\zeta
_{m(a)}\equiv {\kappa _{m(a),c}}/{\kappa _{m(a)}}$ and the system
cooperativity as $C_{g}={4g^{2}}/{\kappa _{m}\kappa _{a}}$. Fig. \ref{f3}(a)
illustrates how the cooperativity affects the transduction efficiency of the
EO system under different coupling ratios.

To further describe the transduction performance of
this system, we introduce the quantum capacity of a single-mode Gaussian
channel, which is lower bounded by the following expression \cite{45,95,96}%
\begin{equation}
Q_{\mathrm{LB}} =
\begin{cases}
\max\Bigl\{0, \log_2\bigl|\frac{\eta}{1-\eta}\bigr| - g(n_e)\Bigr\}\text{,}
& \eta \neq 1\text{,} \\[8pt]
\max\Bigl\{0, \log_2\Bigl(\frac{2}{e\sigma^{2}}\Bigr)\Bigr\}\text{,} & \eta
= 1\text{,}%
\end{cases}
\label{e8}
\end{equation}%
with the added noise $n_{e}$ (Detailed descriptions will be presented in
Sec. IV) and the function $g(n_{e})=(n_{e}+1)\log _{2}(n_{e}+1)-n_{e}\log
_{2}n_{e}$. Here, $\sigma^{2}$ is the noise variance of the random displacement channel. In the low-temperature
limit, this lower bound becomes the exact quantum capacity of the channel.

The probe is assumed to be in the vacuum state. Fig. \ref{f3}(b) shows how $%
Q_{\mathrm{LB}}$ varies with the cooperativity under different coupling
ratios and working temperatures. It can be seen that when the operating
temperature is around $T\sim 0.01\,\mathrm{K}$, the quantum capacity of the
transduction channel is almost identical to that in the low-temperature
limit. This is because the thermal excitation number at microwave frequency $%
\omega _{m}\sim 10\,\mathrm{GHz}$ is $10^{-20}$, which is negligible. As the temperature rises to $0.3\,\mathrm{K}$,
the negative impact of thermal noise on $Q_{\mathrm{LB}}$ increases. However, recent experiments have demonstrated efficient cooling and
several mK temperature can be achieved routinely \cite{88,89,90}.

Comparing Fig. \ref{f3}(a) and \ref{f3}(b), it can be observed that even for
relatively high coupling ratios, the transduction efficiency $\eta$ must
exceed $0.5$ to overcome the threshold for positive quantum capacity.
However, the state-of-the-art $C_{g}$ still falls below the level indicated by
the black arrow, highlighting the need for improved transduction schemes.

\section{Full Representation of the Transducer Model}

\begin{figure}[tbh]
\centering \includegraphics[width=8cm]{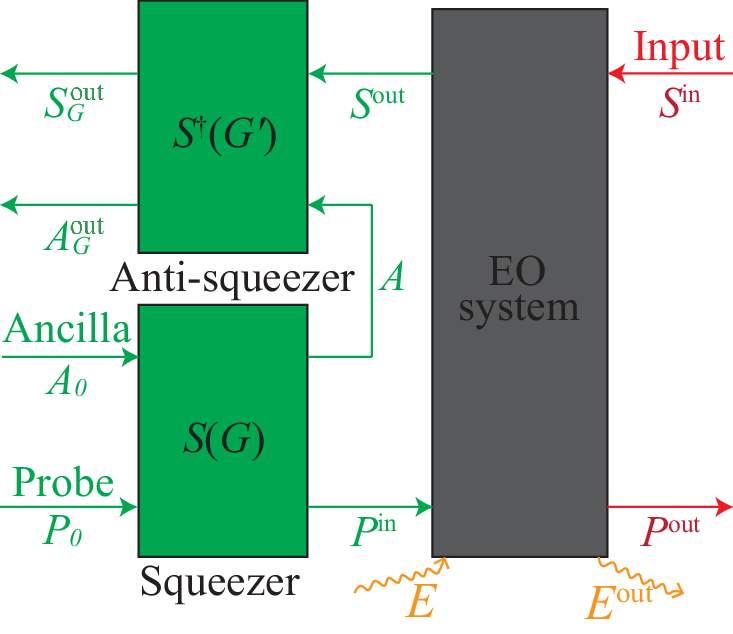}
\caption{(Color online) Schematic diagram of the entanglement-assisted
transducer. It consists of an EO system situated between a squeezer $S(G)$
and an anti-squeezer $S^{\dag }(G^{\prime })$, with squeezer strengths $G$
and $G^{\prime }$, respectively. The probe signal $P_{0}$ and ancilla signal
$A_{0}$, operating in the microwave frequency range (green), are entangled
through the two-mode squeezer $S(G)$, with their outputs denoted by $P^{%
\mathrm{in}}$ and $A$, respectively. Subsequently, $P^{\mathrm{in}}$, along
with the input signal $S^{\mathrm{in}}$ operating in the optical frequency
range (red), is sent to the input port of the EO system. Here, $E^{\mathrm{%
out}}$ is the output of the loss port $E$. After the transduction, the
output $S^{\mathrm{out}}$ and $A$ are subsequently processed by the
anti-squeezer $S^{\dag }(G^{\prime })$, resulting in final outputs $S_{G}^{%
\mathrm{out}}$ and $A_{G}^{\mathrm{out}}$, respectively.}
\label{f2}
\end{figure}

\begin{figure*}[tbh]
\centering \includegraphics[width=18cm]{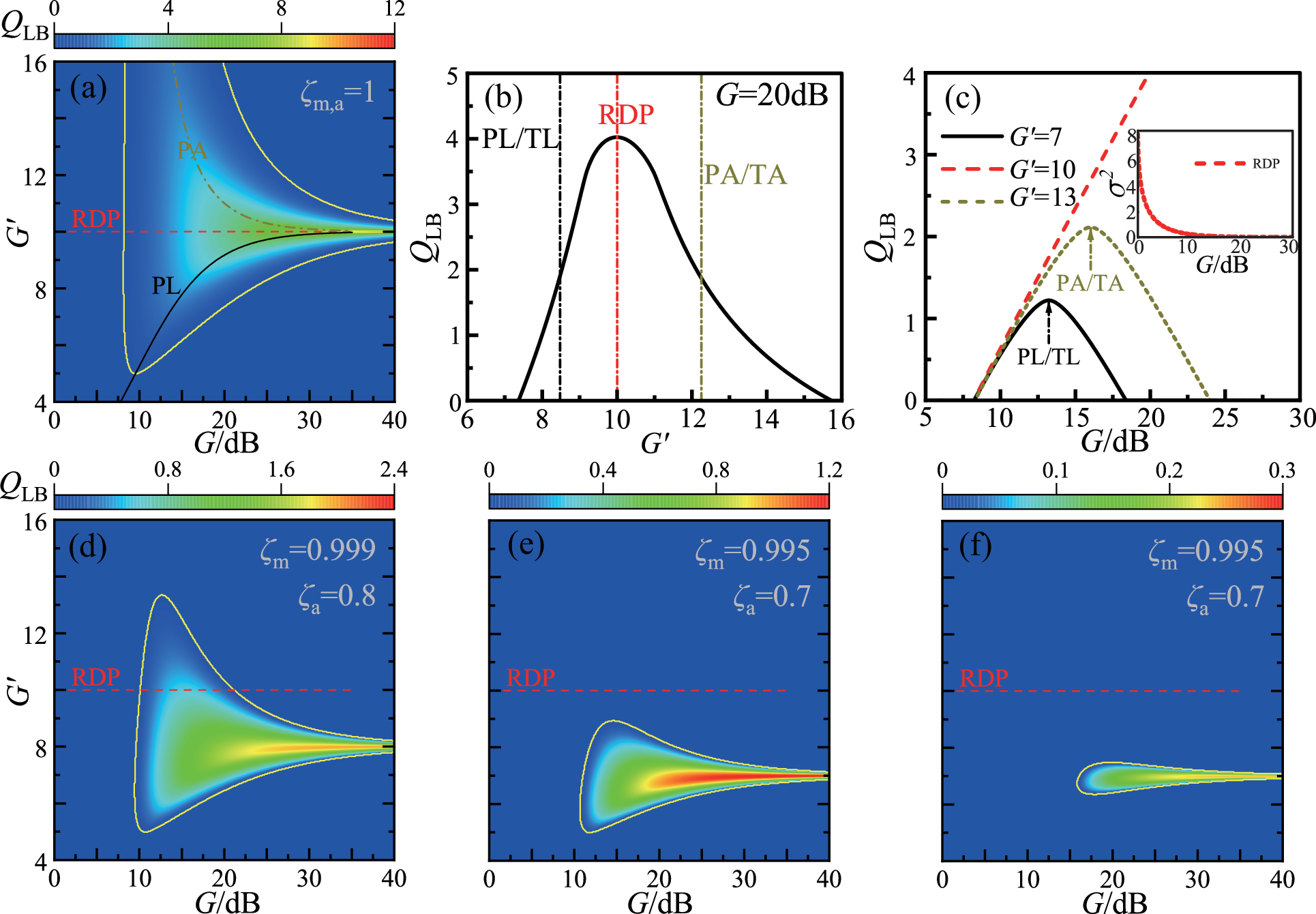}
\caption{(Color online) Transduction performance of the entanglement-assisted transducer. For $\protect\zeta _{m} =\zeta _{a}=1$, (a)-(c) clearly demonstrate the relationship between $Q_{\mathrm{LB}}$ and the squeezer strengths $G$ (expressed in dB as $10\times \log _{10}G$)
and $G^{\prime }$. Here, (b) is plotted with $G$
fixed at $20$ dB, while the three curves in (c) correspond to the fixed
values of $G^{\prime }$ at $7$, $10$, and $13$, respectively. The inset in
(c) displays the noise variance of the RDP channel $\sigma^{2}$ as a
function of $G$. For various non-unity coupling ratios ($\protect\zeta _{m,a}\neq1$), (d)-(f) plot $Q_{\mathrm{LB}}$ versus the squeezer strengths $G$ and $G'$. Here, (d) and (e) correspond to the
transduction channel performance in the low-temperature limit, whereas (f)
corresponds to the case at $0.3$ K. In (a) and (d)-(f), the yellow curves
represent the $Q_{\mathrm{LB}}>0$ boundary.  The other parameters are chosen
as $\protect\eta =0.1$ and $\protect\omega _{m}=2%
\protect\pi \times 10$ GHz.}
\label{f5}
\end{figure*}

As illustrated in Fig. \ref{f2}, the intra-band entanglement-assisted transducer
consists of a nonlinear EO system situated between a squeezer $S(G)$ and an
anti-squeezer $S^{\dag }(G^{\prime })$. The system involves three input
fields: the probe $P_{0}$, the ancilla $A_{0}$, and the input signal $S^{%
\mathrm{in}}$. Here, the probe $P_{0}$ and the ancilla $A_{0}$ are both in
the vacuum states and entangled through a two-mode squeezing interaction
generated by $S(G)$. Subsequently, $S^{\mathrm{in}}$ and $P^{\mathrm{in}}$
are sent to the input ports of the EO system. Finally, the output $S^{%
\mathrm{out}}$ and $A$ are anti-squeezed by $S^{\dag }(G^{\prime })$,
resulting in the final converted output $S_{G}^{\mathrm{out}}$ in the
microwave frequency. Notably, the final output $S_{G}^{\mathrm{out}}$ is the
sum of the three initial signals and the noise from the loss port $E$, where each component
can be altered by adjusting the squeezing strengths $G$ and $G^{\prime }$.

We first assume the
system is on resonant, thus, $\omega $ is omitted in the
subsequent discussion. The non-resonant situation will be
addressed later in the text. For the squeezers $S(G)$ and $S^{\dag
}(G^{\prime })$, the corresponding input-output relations are given by%
\begin{equation}
\varepsilon _{P^{\mathrm{in}}}=\sqrt{G}\varepsilon _{P_{0}}+\sqrt{G-1}%
\varepsilon _{A_{0}}^{\dag }\text{,}
\end{equation}%
\begin{equation}
\varepsilon _{A}^{\dag }=\sqrt{G-1}\varepsilon _{P_{0}}+\sqrt{G}\varepsilon
_{A_{0}}^{\dag }\text{,}  \label{e9}
\end{equation}%
and%
\begin{equation}
\varepsilon _{S_{G}^{\mathrm{out}}}=\sqrt{G^{\prime }}\varepsilon _{S^{%
\mathrm{out}}}-\sqrt{G^{\prime }-1}\varepsilon _{A}^{\dag }\text{.}
\label{e10}
\end{equation}%
Combining with Eq. (\ref{e4}), we can obtain the full input-output relation
for the entanglement-assisted transducer%
\begin{eqnarray}
\varepsilon _{S_{G}^{\mathrm{out}}} &=&\sqrt{\eta G^{\prime }}\varepsilon
_{S^{\mathrm{in}}}+\sqrt{\kappa _{E}G^{\prime }}\varepsilon _{E}  \notag \\
&&+[\sqrt{GG^{\prime }\kappa _{P}}-\sqrt{(G-1)(G^{\prime }-1)}]\varepsilon
_{P_{0}}  \label{e12} \\
&&+[\sqrt{(G-1)G^{\prime }\kappa _{P}}-\sqrt{G(G^{\prime }-1)}]\varepsilon
_{A_{0}}^{\dag }\text{,}  \notag
\end{eqnarray}%
where the four terms represent the contributions from the input signal,
loss, probe, and ancilla ports, respectively. Interestingly, we see the transduction efficiency is enhanced by $G^\prime$. In the following section, we will show that the squeezing
strengths can also be tuned to realize different quantum transduction channels.

\section{Characterization of intra-band entanglement-assisted transduction Channels}

\subsection{Channel classification}

To facilitate a clear analysis of transduction channels, we assume that $%
\zeta _{m} =\zeta _{a}=1$, eliminating the loss port $E$ term in Eq. (\ref{e12}).
Consequently, the final output $\varepsilon _{S_{G}^{\mathrm{out}}}$ is
determined by the three input ports and is governed by the squeezer
strengths $G$ and $G^{\prime }$ as well as the transduction
efficiency $\eta $. Now, we can basically classify the transduction channels
based on the enhanced transduction efficiency $G^{\prime }\eta $: (i)
generalized loss (GL) channel for $G^{\prime }\eta <1$; (ii) generalized
amplification (GA) channel for $G^{\prime }\eta >1$; and (iii) random
displacement (RDP) channel for $G^{\prime }\eta =1$.

In order to rigorously establish the input-output relations for these
transduction channels, we define the total noise operator $\varepsilon _{e}$%
, which satisfies the canonical commutation relation $[\varepsilon
_{e},\varepsilon _{e}^{\dag }]=1$. This requirement specifies that, for the
GL channel, $\varepsilon _{e}$ takes the form%
\begin{eqnarray}
&&\varepsilon _{e}=  \notag \\
&&\frac{1}{\sqrt{1-\eta G^{\prime }}}\Biggl\{\Bigl[\sqrt{GG^{\prime }\kappa
_{P}}-\sqrt{(G-1)(G^{\prime }-1)}\Bigr]\varepsilon _{P_{0}}  \notag \\
&&\quad +\Bigl[\sqrt{(G-1)G^{\prime }\kappa _{P}}-\sqrt{G(G^{\prime }-1)}%
\Bigr]\varepsilon _{A_{0}}^{\dag }\Biggr\}\text{,}
\end{eqnarray}%
while for the GA channel, the corresponding noise operator $\varepsilon
_{e}^{\dag }$ is given by
\begin{eqnarray}
&&\varepsilon _{e}^{\dag }=  \notag \\
&&\frac{1}{\sqrt{\eta G^{\prime }-1}}\Biggl\{\Bigl[\sqrt{GG^{\prime }\kappa
_{P}}-\sqrt{(G-1)(G^{\prime }-1)}\Bigr]\varepsilon _{P_{0}}  \notag \\
&&\quad +\Bigl[\sqrt{(G-1)G^{\prime }\kappa _{P}}-\sqrt{G(G^{\prime }-1)}%
\Bigr]\varepsilon _{A_{0}}^{\dag }\Biggr\}\text{.}
\end{eqnarray}%
Thus, the transduction channels can be written in canonical form: (i) the GL
channel%
\begin{equation}
\varepsilon _{S_{G}^{\mathrm{out}}}=\sqrt{\eta G^{\prime }}\varepsilon _{S^{%
\mathrm{in}}}+\sqrt{1-\eta G^{\prime }}\varepsilon _{e}\text{;}
\end{equation}%
and (ii) the GA channel%
\begin{equation}
\varepsilon _{S_{G}^{\mathrm{out}}}=\sqrt{\eta G^{\prime }}\varepsilon _{S^{%
\mathrm{in}}}+\sqrt{\eta G^{\prime }-1}\varepsilon _{e}^{\dag }\text{.}
\end{equation}%
The RDP channel exhibits an asymptotic behavior that bridges the GL and GA
channels as $\eta G^{\prime }\rightarrow 1$, and its input-output relation
is given by
\begin{eqnarray}
\varepsilon _{S_{G}^{\mathrm{out}}} &=&\varepsilon _{S^{\mathrm{in}}}+\sqrt{%
\frac{1}{\eta }}\Bigl[\sqrt{G\kappa _{P}}-\sqrt{(G-1)(1-\eta )}\Bigr]%
\varepsilon _{P_{0}}  \notag  \label{rdp_eq} \\
&&-\sqrt{\frac{1}{\eta }}\Bigl[\sqrt{(G-1)\kappa _{P}}-\sqrt{G(1-\eta )}%
\Bigr]\varepsilon _{A_{0}}^{\dag }\text{.}
\end{eqnarray}

Notably, the GL and GA channels each have a unique case. Specifically, for
the GL channel, when $G^{\prime
}=G/[G(1-\kappa _{P})+\kappa _{P}]$, the $\varepsilon _{A_{0}}^{\dag }$ term in the total
noise operator $\varepsilon _{e}$ is eliminated. The final output simplifies to
\begin{equation}
\varepsilon _{S_{G}^{\mathrm{out}}}=\sqrt{\frac{G\eta }{G(1-\kappa
_{P})+\kappa _{P}}}\varepsilon _{S^{\mathrm{in}}}+\sqrt{\frac{\kappa _{P}}{%
G(1-\kappa _{P})+\kappa _{P}}}\varepsilon _{P_{0}}\text{,}
\label{noancilla_eq}
\end{equation}%
which corresponds to the pure-loss (PL) channel in the low-temperature limit
or the thermal-loss (TL) channel at non-zero temperature. This is exactly
the transduction channel that Ref. \cite{87} discussed. Moreover, for GA
channel, when the $\varepsilon _{P_{0}}$ term in the total noise operator $%
\varepsilon _{e}^{\dag }$ is eliminated, i.e., $G^{\prime
}=(G-1)/[G(1-\kappa _{P})-1]$, the output simplifies to
\begin{equation}
\varepsilon _{S_{G}^{\mathrm{out}}}=\sqrt{\frac{(G-1)\eta }{G(1-\kappa
_{P})-1}}\varepsilon _{S^{\mathrm{in}}}+\sqrt{\frac{\kappa _{P}}{G(1-\kappa
_{P})-1}}\varepsilon _{A_{0}}^{\dag }\text{,}  \label{noprobe_eq}
\end{equation}%
which corresponds to the pure-amplification (PA) channel in the
low-temperature limit and the thermal-amplification (TA) channel at non-zero
temperature.

\subsection{Quantum capacity versus squeezer strengths}

Based on Eq. (\ref{e8}), we can obtain the quantum capacity lower bound of
the entanglement-assisted transducer by substituting the
transduction efficiency $\eta$ and the EO system's added noise $n_{e}$ with
the enhanced transduction efficiency $G^{\prime }\eta $ and its
corresponding noise $N_{e}$, respectively. Here, $N_{e}$ is determined by the new
bosonic mode $N_{e}=\langle \varepsilon _{e}^{\dag }\varepsilon _{e}\rangle $%
. In this subsection, we fix $\eta
=0.1$. Fig. \ref{f5}(a) shows how $Q_{\mathrm{LB}}$ varies with respect to
squeezer strengths $G$ and $G^{\prime }$. Remarkably, a large parameter regime for positive quantum capacity can be achieved, as enclosed by the yellow curve. Here, the region above the
red-dashed line, which represents the RDP channel, corresponds to the GA
channel, while the region below corresponds to the GL channel. It is
noticeable that, for a fixed $G$, $Q_{\mathrm{LB}}$ reaches its maximum
value at $G^{\prime }=1/\eta $ (RDP channel). Moreover, the $Q_{\mathrm{LB}%
}>0$ boundary exhibits a convex shape along the curves corresponding to the
PL/TL and PA/TA channels. This indicates that, for a fixed $G^{\prime }$,
the local maximum values of $Q_{\mathrm{LB}}$ are located on these curves,
highlighting their significance in optimizing
the transduction channel. As $G$ increases, the range of $G^{\prime }$ where
$Q_{\mathrm{LB}}>0$ becomes narrower. However, the maximum value of $Q_{%
\mathrm{LB}}$ increases at the same time, reflecting the effective
suppression of noise in the RDP channel with larger $G$.

Next, we further investigate the respective relationships of $Q_{\mathrm{LB}}$ with $G^{\prime }$ and $G$. As shown in
Fig. \ref{f5}(b), for a fixed $G$, the region to the left of the red
vertical line, which corresponds to the RDP channel, represents the GL
channel, while the region to the right represents the GA channel. The two
vertical lines within the GL and GA channels correspond to the PL/TL and
PA/TA channels, respectively. Consistent with the situation in Fig. \ref{f5}%
(a), the $Q_{\mathrm{LB}}$ of the GL channel increases with $G^{\prime }$,
reaching a maximum at the RDP channel, and then decreases as the channel
transitions into the GA regime with larger $G^{\prime }$. This
trend implies that for the GL channel, the amplification of the input signal
$\varepsilon _{S^{\mathrm{in}}}$ by $G^{\prime }$ is prominent, whereas in
the GA channel, the noise amplification gradually becomes the dominant
effect as $G^{\prime }$ increases. Moreover, for a fixed $G^{\prime }$, Fig. %
\ref{f5}(c) illustrates how the quantum capacities of the GL, RDP, and GA
channels vary as a function of $G$. In particular, the GL (GA) channel
reaches its maximum $Q_{\mathrm{LB}}$ value when it operates as a PL/TL
(PA/TA) channel, which is indicated by the arrow on the curve of $G^{\prime
}=7$ ($G^{\prime }=13$). At $G^{\prime }=10$, we get the RDP channel with a unit transmissivity, and its added noise is linearly suppressed as $G$ increases (see the
inset of Fig. \ref{f5}(c)). Consequently, the system's quantum capacity
exhibits a log-linear dependence on $G$.

When the loss port $E$ is taken into account, the noise term $\varepsilon
_{E}$ in Eq. (\ref{e12}) adversely affects the quantum capacity $Q_{\mathrm{%
LB}}$ of the transduction channel. In fact, in the final output $\varepsilon
_{S_{G}^{\mathrm{out}}}$, the amplification provided by $S^{\dag }(G^{\prime
})$ boosts not only the input signal $\varepsilon _{S^{\mathrm{in}}}$ but
also the noise operator $\varepsilon _{E}$. Consequently, at higher $%
G^{\prime }$ values the channel becomes increasingly sensitive to $%
\varepsilon _{E}$, resulting in a more pronounced reduction in $Q_{\mathrm{LB%
}}$ and a downward shift of its maximum, which appears in the lower region
of the RDP channel, as depicted in Fig. \ref{f5}(d) and \ref{f5}(e).
Correspondingly, the $Q_{\mathrm{LB}}>0$ boundary contracts more noticeably
as $G^{\prime }$ increases. At non-zero operating temperature, the thermal
noise further diminishes the overall $Q_{\mathrm{LB}}$ of the transduction
channel, as illustrated in Fig. \ref{f5}(f).

\section{Comparative Analysis under the Non-Resonant Condition}

In this section, we compare the quantum capacity of the
transducer with and without entanglement assistance under the
non-resonant condition. Here, we focus on the PL channel in the
low-temperature limit. Serving as auxiliary resources, the two squeezers play
complementary roles in the entanglement-assisted transducer. On one hand,
the squeezer strength $G^{\prime }$ can effectively enhance the
transduction efficiency $\eta $. On the other hand, as $G$ increases, the
noise introduced by the probe $P_{0}$ can be effectively suppressed.
Consequently, the PL channel implemented by the entanglement-assisted
transducer shows a significant improvement in the quantum capacity compared
to the bare EO system. Additionally, the transduction
bandwidth corresponding to the positive quantum capacity is also notably
increased.

\begin{figure}[tbh]
\centering \includegraphics[width=\columnwidth]{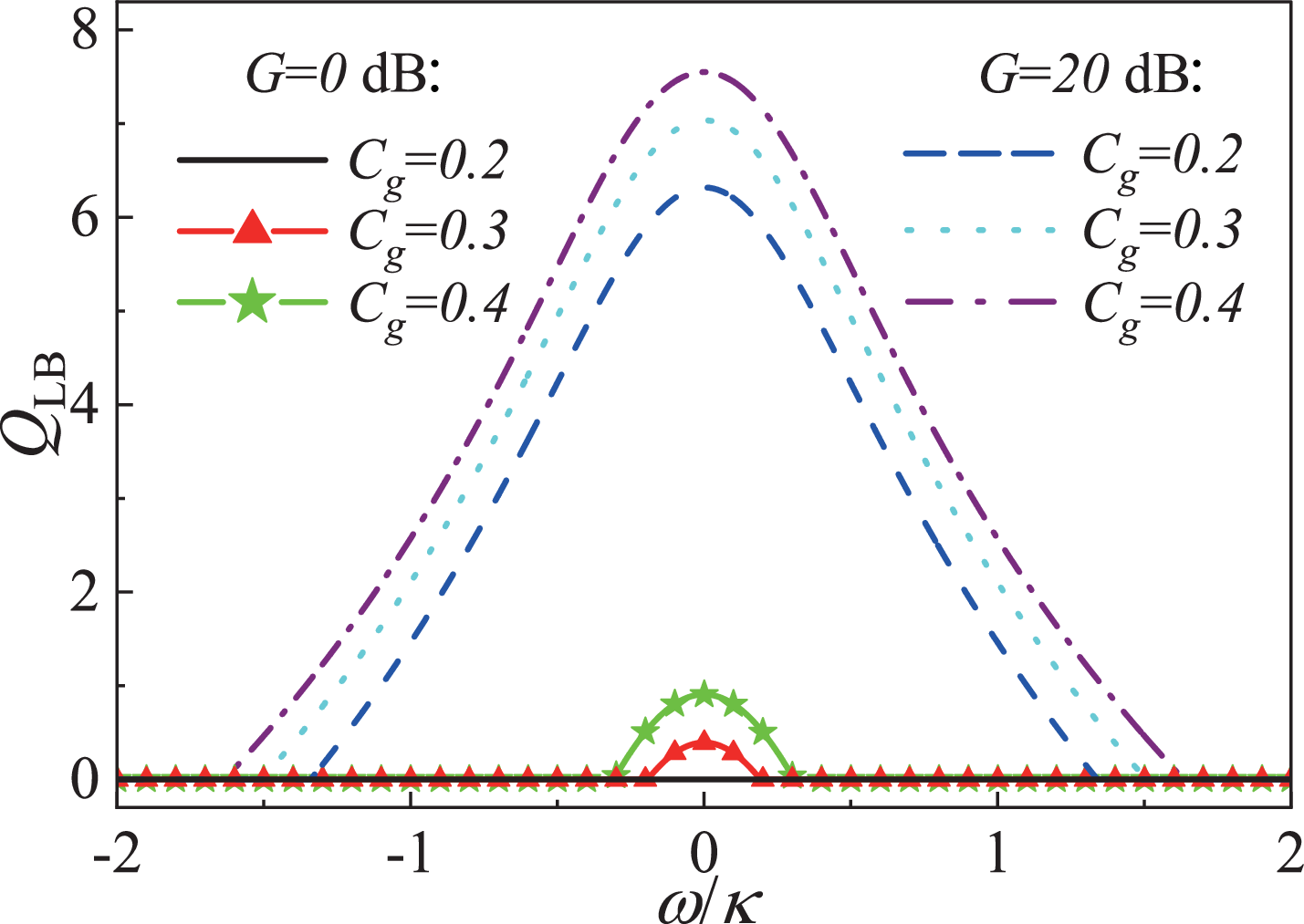}
\caption{(Color online) Quantum capacity $Q_{\mathrm{LB}}$ distribution in
the frequency domain for both the entanglement-assisted transducer and the
bare EO system at the low-temperature limit. The coupling ratios are chosen
as $\protect\zeta _{m} =0.999$ and $\protect\zeta _{a} =0.8$.}
\label{f0}
\end{figure}

Without loss of generality, we set $\kappa _{m}=\kappa _{a}=\kappa$.
According to the transduction efficiency of the EO system under the
non-resonant condition Eq. (\ref{eta_intrinsic_eo}), the frequency-domain
distributions of the quantum capacity for the PL channel implemented by both
the bare EO system and the entanglement-assisted transducer are illustrated
in Fig. \ref{f0}. In the absence of entanglement assistance ($G=0$ dB), the
bare EO system fails to achieve a positive quantum capacity when $C_{g}=0.2$%
, where $\eta(\omega=0)<0.5$. In contrast, the entanglement-assisted
transducer can realize a PL channel with a high-bandwidth positive quantum
capacity. As the cooperativity $C_{g}$ increases, the bare EO system begins
to exhibit positive quantum capacity near the resonant frequency, while the
entanglement-assisted transducer attains a higher quantum capacity over a
much broader bandwidth. Notably, for systems with lower $C_{g}$,
entanglement assistance yields a more pronounced enhancement in the
effective bandwidth, thereby enhancing the transducer's practical
applicability.

\section{Conclusion}

In conclusion, we study an entanglement-assisted quantum transducer based on
a cavity EO system to overcome the high threshold required for achieving
positive quantum capacity. By introducing an assist mode and two squeezers,
the transduction efficiency is significantly enhanced, greatly lowering the
threshold for the positive quantum capacity compared to the bare EO system.
We presented a detailed analysis of the transducer, exploring the three
types of transduction channels achievable through the adjustment of the
squeezing strength of the two squeezers. Moreover, we examine how a broad
variation in the two squeezing strengths influences the quantum capacities of
different transduction channels, clarifying the conditions needed to fully
optimize the transducer's
performance. Additionally, the entanglement-assisted transducer has
a marked improvement in the bandwidth for achieving high quantum capacity. This advancement is crucial for
constructing high-bandwidth DQTs, while ensuring high-fidelity signal
transduction. Our study provides a full theoretical framework for analyzing intra-band entanglement-assisted quantum
transduction scheme, which unlocks more potentials of its application in future quantum technologies.

\begin{acknowledgements}
This work was supported the start-ups from Xi'an Jiaotong University (Grant No. 11301224010717).
\end{acknowledgements}

\bibliography{sub.bib}

\end{document}